%% file: eprint_dpf2015_nova.tex
\documentclass[12pt]{article}
\usepackage{graphicx}
\usepackage{amsmath}

\newcommand\pubnumber{DPF2015-218}
\newcommand\pubdate{\today}

\def\umn{School of Physics and Astronamy\\
University of Minnesota, MN 55455, USA}

\def\dedx{d$E$/d$x$}

\def\Title#1{\begin{center} {\Large #1 } \end{center}}
\def\Author#1{\begin{center}{ \sc #1} \end{center}}
\def\Address#1{\begin{center}{ \it #1} \end{center}}

\newcommand\pubblock{\rightline{\begin{tabular}{l} \pubnumber\\
         \pubdate  \end{tabular}}}
\newenvironment{Abstract}{\begin{quotation}  }{\end{quotation}}
\newenvironment{Presented}{\begin{quotation} \begin{center}
             PRESENTED AT\end{center}\bigskip
      \begin{center}\begin{large}}{\end{large}\end{center} \end{quotation}}

\input econfmacros.tex
\begin{document}
\begin{titlepage}
\pubblock

\vfill
\Title{ First Results of $\nu_e$ Appearance Analysis and Electron Neutrino Identification at NOvA}
\vfill
\Author{ Jianming Bian \\
(for the NOvA Collaboration)}
\Address{\umn}
\vfill
\begin{Abstract}
NOvA is a long-baseline accelerator-based neutrino oscillation experiment that is optimized for $\nu_\mu\to\nu_e$ measurements. It uses the upgraded NuMI beam from Fermilab and measures electron-neutrino appearance and muon-neutrino disappearance at its Far Detector in Ash River, Minnesota. The $\nu_e$ appearance analysis at NOvA aims to resolve the neutrino mass hierarchy problem and to constrain the CP-violating phase.
The first data set of $2.74\times10^{20}$ protons on target (POT) equivalent exposure taken by NOvA has been analyzed. The first measurement of electron-neutrino appearance in NOvA provides solid evidence of $\nu_\mu\to\nu_e$ oscillation with the NuMI beam line. Electron-neutrino identification is the key ingredient for the $\nu_e$ appearance analysis. The electron-identification algorithm used to produce the primary results presented here compares 3-D shower-energy profiles with Monte Carlo prototypes to construct likelihoods for each particle hypothesis. Particle likelihoods, among other event-topology variables, are used as inputs to an Artificial Neural Network for the final electron-neutrino identification. The design and implementation of this algorithm is also presented.

%We developed a new method to precisely describe transverse and longitudinal shower energy profile for particle identification at NO?A. Likelihoods under different particle hypotheses based on detailed shower transverse and longitudinal developments are calculated.Using these likelihoods, a neural network has been trained and applied to identification of electrons for analysis of $\nu_e$ events.

\end{Abstract}
\vfill
\begin{Presented}
DPF 2015\\
The Meeting of the American Physical Society\\
Division of Particles and Fields\\
Ann Arbor, Michigan, August 4--8, 2015\\
\end{Presented}
\vfill
\end{titlepage}
\def\thefootnote{\fnsymbol{footnote}}
\setcounter{footnote}{0}

\section{Introduction}

NOvA (NuMI Off-Axis $\nu_e$ Appearance Experiment)  is a neutrino experiment optimized to observe the oscillation of muon neutrinos to electron-neutrinos \cite{ref:nova}. NOvA uses a 14-kt liquid scintillator Far Detector (FD) in Ash river, Minnesota to detect the oscillated NuMI (Neutrinos at the Main Injector) muon neutrino beam produced 810 km away at Fermilab \cite{ref:numi}. The NOvA baseline is the longest in operation, which maximizes the matter effect and allows a measurement of the neutrino mass ordering. NOvA is equipped with a  0.3-kt functionally identical Near Detector (ND) located at Fermilab to measure unoscillated beam neutrinos and estimate backgrounds at the FD. Both detectors are located 14 mrad off-axis to receive a narrow-band neutrino energy spectrum near the energy of the $\nu_\mu\to\nu_e$ oscillation maximum range ($\sim$2 GeV), enhancing the $\nu_\mu\to\nu_e$ oscillation signal in the FD while reducing neutral current and beam $\nu_e$ backgrounds from high-energy, unoscillated beam neutrinos.

NOvA's detectors consist of plastic (PVC) extrusions filled with liquid scintillator, with wavelength shifting fibers (WLS) connected to avalanche photodiodes (APDs). The dimensions of the detector cells are 6 cm $\times$ 4 cm, with each cell extending the full width or height of the detector, 15.6 m in the FD and 4.1 m in the ND. Extrusions are assembled in alternating layers of vertical and horizontal extrusions plane, so 3-D hit information is available for clustering and particle identification. Each plane (cell width) of the detectors is  just 0.15 radiation lengths ($X_0$). This level of granularity helps greatly to separate electrons from $\pi^0$ backgrounds \cite{ref:det1,ref:det2,ref:det3}.

NO$\nu$A measures $\nu_e$ ($\bar{\nu_e}$) appearance probability and $\nu_\mu$ ($\bar{\nu_\mu}$) disappearance probability with neutrino and anti-neutrino beams. The $\nu_e$ ($\bar{\nu_e}$) appearance experiment investigates (1) the neutrino mass hierarchy, (2) the CP-violation phase in the neutrino sector and (3) the $\theta_{23}$ octant (whether $\theta_{23}>$ or $ <45^\circ$). For the first $\nu_e$ analysis, we use 2.74$\times 10^{20}$ POT equivalent exposure (1/13th of the overall planned exposure) of neutrino running to measure the probability of $\nu_\mu\to\nu_e$. Although there is no information from $\bar{\nu_\mu}\to\bar{\nu_e}$ in the first analysis, in a favorable $\delta_{CP}$ region around $3\pi/2$, there is some sensitivity to discriminate the mass hierarchy.

The first $\nu_e$ appearance measurement is a cut-and-count analysis \cite{ref:nuefa}.  A $\nu_e$ event  is identified in charged current (CC) interactions where the electron-neutrino converts into an electron.
The $\nu_e$ analysis at NOvA makes use of two electron event identification algorithms (EID). The primary EID algorithm is LID, which is an artificial neural network using 3-D shower-energy profile based likelihoods for particle hypotheses. The second EID is LEM, which matches trial events to an enormous Monte Carlo library.  Implementation of the LID algorithm is  described in Section~\ref{sec:lid}; detail about the LEM algorithm can be found in Ref.~\cite{Backhouse:2015xva}.

\section{\bf Electron Neutrino Identification (LID)}\label{sec:lid}

The basic idea of LID is to use the shower-energy profile to separate electrons from muons, $\pi^0$s, and other hadrons.  Different particles have very different energy-depositions in the detector. For example, the electron deposits energy through ionization in the first few planes then starts a shower; the photon is a shower that follows a gap in the first few planes;  and the muon registers as a long minimum ionizing particle (MIP) track. This makes it possible to identify particles by comparison of shower/track shapes with different particle hypotheses.

%Figure~\ref{fig:avglongdedx} and \ref{fig:avgtransdedx} show average longitudinal and transverse \dedx\ for $e$, $\gamma$, $\mu$ and $\pi^0$, which is a simple metric of particle energy profile.\\
%Figure~\ref{fig:avglongdedx} and \ref{fig:avgtransdedx} show average longitudinal and transverse \dedx\ for $e$, $\gamma$, $\mu$ and $\pi^0$, which is a simple metric of particle energy profile.\\

%\begin{figure}[htb]
%\centering
%\includegraphics[height=3.5in]{eventtopology.eps}
%\caption{Event display for $e$, $\gamma$ and $\mu$.}
%\label{fig:evd}
%\end{figure}

%\begin{figure}[htb]
%\centering
%\includegraphics[height=3.5in]{avglongdedx.eps}
%\caption{Event display for $e$, $\gamma$ and $\mu$.}
%\label{fig:avglongdedx}
%\end{figure}

%\begin{figure}[htbp]
%\begin{center}
%\epsfig{file=eventtopology.eps, width = 12 cm, height = 8 cm} \caption{Event display for $e$, $\gamma$ and $\mu$.}
%\label{fig:evd}
%\end{center}
%\end{figure}

To precisely identify particles, we cluster showers from reconstructed collections of cell hits with a start point and direction (prong), and then compare the measured longitudinal and transverse \dedx\ of candidate showers with the expected distributions found in MC samples for each particle hypothesis to obtain likelihoods. We perform this comparison plane-by-plane in the longitudinal direction and cell-by-cell in the transverse direction. In this way we can make use of all energy-profile information in a shower.  Using these likelihoods, a neural network has been trained and applied to the identification of $\nu_e$ CC events.

\subsection{\bf $\nu_e$ event reconstruction}

The $\nu_e$ event reconstruction begins with clustering hits by space-time coincidence to separate beam events from cosmic rays in a trigger window. This procedure can collect together hits from a single neutrino interaction (slice). The slices then serve as the foundation for all later reconstruction stages \cite{ref:reco1}. Next, a modified Hough transform is applied to identify prominent straight-line features in a slice. Then the lines are tuned in an iterative procedure until they converge to the interaction vertex of that slice. Prongs are then reconstructed based on distances from hits to the lines associated with each of the particle that paths emanating from the reconstructed vertex \cite{ref:reco2}-\cite{Niner:2015aya}.

\subsection{\bf Shower clustering}

We define the shower core based on the prong direction provided by the prong cluster, then collect signal hits in a column around this core. To reduce the contamination from the hadronic interaction around the vertex, we require the radius to be twice the cell width for the first 8 planes from the start point of the shower and $20$ times the cell width for other planes. In this way, there is good efficiency for including all hits caused by an electron in the shower.

%\begin{figure}[htb]
%\centering
%\includegraphics[height=1.8in]{reclustering.eps}
%\caption{Shower clustering}
%\label{fig:reclustering}
%\end{figure}

\subsection{\bf Cell energy deconvolution}

Daughter photons from a high-momentum background $\pi^0$ and a shower in a neutrino event could partially overlap. To handle this, we perform a de-convolution when we detect overlapping showers based on total energy or distance to the cores. For a cell associated with more than one shower, we determine the energy associated with the $i$-th shower as follows:

$$\displaystyle E_i^{cell}=\frac{PE^{cell}}{a_i}\cdot\frac{(PE_i^{shower}P_i/a_i)}{\sum_{i}(PE_j^{shower}P_j/a_j)},$$
$$P_j=\exp(-D_i/\lambda),$$

\noindent where $E_i^{cell}$ is energy in a cell belonging to the shower, $PE^{shower}_{i}$ is the total number of photoelectrons (PE) in the shower, $PE^{cell}$ is the total PE in a cell, $a_i$ is a factor that scales from PE to GeV and corrects for attenuation based on distance to readout, $D_i$ is the distance from the cell to the core of the shower, and $\lambda$ is a constant for the shower lateral profile. Here we assume the transverse \dedx\ as a function of distance to the shower core behaves as follows:

$$d{\rm E}/d{\rm x} (x) = AE\exp(-x/\lambda),$$

\noindent where A is the normalization constant, $E$ is the shower energy and $x$ is the distance to the shower core. By fitting the transverse shower energy profile of simulated electrons from $\nu_e$ we determine $\lambda$ to be 3.05 cm.

%, as shown in Figure~\ref{fig:fitavgtransdedx}.
%
%\begin{figure}[htb]
%\centering
%\includegraphics[height=1.8in]{fitavgtransdedx.eps}
%\caption{(average transverse dE/dx)/(shower energy) vs. distance for electron selected from $\nu_e$ MC and the fit.}
%\label{fig:fitavgtransdedx}
%\end{figure}

%and $R_M$ is the Moliere Radius of the scintillator
% Figure~\ref{fig:mergedpi0} displays this case in a 2.0 GeV $\pi^0$ MC event.
%\begin{figure}[htb]
%\centering
%\includegraphics[height=1.8in]{mergedpi0.eps}
%\caption{A $\pi^0$ whose daughter photons are partially overlapped in event display}
%\label{fig:mergedpi0}
%\end{figure}

%Figure~\ref{fig:edeconv} shows an example of energy deconvolution for a $\pi^0$ which has overlapping daughter photons. We can see that after cell energy deconvolution the shower energy for the two photons can be assigned reasonably.
%\begin{figure}[htb]
%\centering
%\includegraphics[height=1.8in]{edeconv.eps}
%\caption{$\pi^0$ daughter photon energy as a function of distance to shower cores, before (black) and after (blue) cell energy deconvolution}
%\label{fig:edeconv}
%\end{figure}

\subsection{\bf $dE$/$dx$ and log likelihood} \label{dedxlikelihood}

For an unidentified particle, we compare the measured \dedx\ with the expected \dedx\ under each particle hypothesis in each plane and transverse cell to construct the probability and likelihood for particle identification. In this way we can describe the 3-D development of a shower in detail.

Using the deconvoluted cell energy, we calculate \dedx\  in the longitudinal and transverse directions. The longitudinal \dedx\  is calculated plane-by-plane. It is defined as the total shower energy in a plane divided by the path length in that plane (the thickness of a plane divided by the cosine of the incident angle of the shower). The transverse \dedx\  is calculated using the following method. (1) A line connecting the start and end point of the shower is constructed. (2) In each plane the cell that the line intersects is considered to be the core of the shower and is assigned a transverse cell index of zero. (3) In a given plane, the next cells out from the core cell in the positive or negative transverse direction are both assigned a transverse cell index of 1,  and so on for transverse cell indices up to 20. (4) For a given transverse cell index, the energies in cells along the entire shower are summed and divided by the total path length to give the total \dedx\ corresponding to that transverse cell index. (5) The average \dedx\  for each transverse cell index is calculated.

\begin{figure}[h]
\begin{center}
\includegraphics[width=12cm]{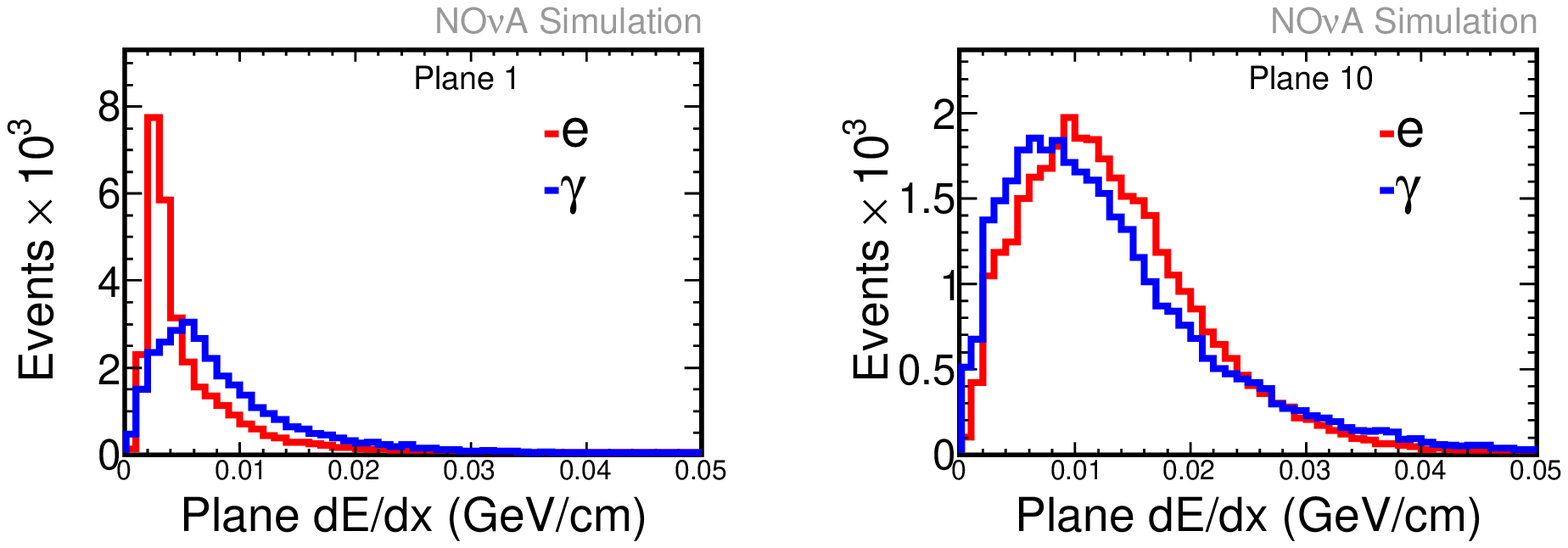}
\caption{Longitudinal d$E$/d$x$ for electrons (red) and photons (blue): (left) plane index = 1; (right) plane index = 10.}\label{fig:eglongdedx}
\includegraphics[width=12cm]{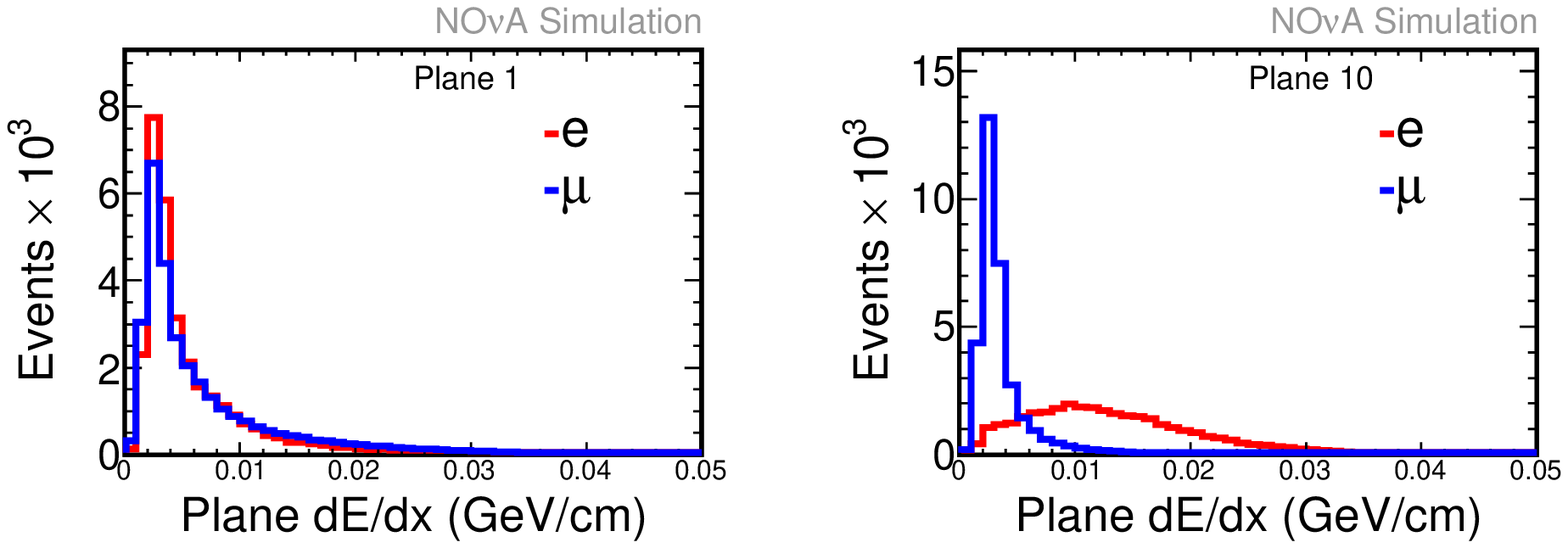}
\caption{Longitudinal d$E$/d$x$ for electrons (red) and muons (blue): (left) plane index = 1; (right) plane index = 10.} \label{fig:emulongdedx}
\end{center}
\end{figure}

By matching the reconstructed shower direction to truth, we select Monte Carlo $e$, $\gamma$, $\mu$, $\pi^0$, $p$, $n$, and $\pi^{\pm}$ showers from neutrino MC events to extract the expected \dedx\ distribution histograms for each plane and each transverse cell index.  To consider energy dependence, we evenly divide the shower energy range 0-5.0~GeV into 10 bins, then obtain \dedx\ histograms in these energy bins. Figures~\ref{fig:eglongdedx} and \ref{fig:emulongdedx} show the expected longitudinal d$E$/d$x$ distributions in different planes for electrons, photons and muons with energy greater than 0.5 GeV. In the second plane (plane index = 1), as shown in Figure~\ref{fig:eglongdedx}(left), the electron has a sharp minimum ionization peak, while the photon has started the EM shower which has a broader distribution in \dedx. This signature provides powerful separation of electrons from $\gamma$'s or $\pi^0$'s. Figure~\ref{fig:emulongdedx} (right) shows that in the 11th plane the electron is a shower whereas the muon still has minimum ionizing deposition, which makes the \dedx\  distributions very different for these two particles.

%\begin{figure}[h]
%\begin{center}
%\includegraphics[width=12cm]{epi0transdedx.eps}
%\caption{Transverse d$E$/d$x$ for electrons (red) and $\pi^0$'s (blue): (left) Transverse cell index = 0; (right) Transverse cell index = 3.}\label{fig:epi0transdedx}
%\end{center}
%\end{figure}

\begin{figure}[]
\begin{center}
\includegraphics[width=10cm]{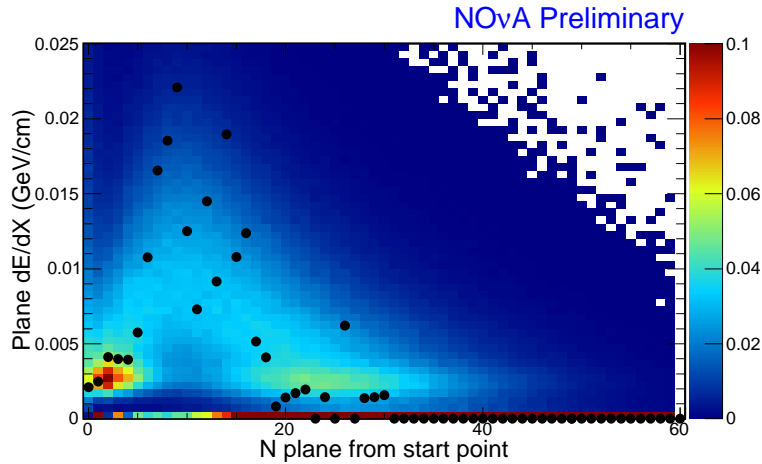}
\caption{Color: p.d.f. for \dedx\ in each plane (electron assumption); Points: measured dE/dx in each plane (example event) }\label{fig:evtlongdedx19578}
\end{center}
\end{figure}

When performing EID on the test sample, we calculate the \dedx\  in each plane and transverse cell index. We then calculate the probability for each type of particle according to the expected \dedx\ histogram for that plane or transverse cell index. We find the bin index in the expected \dedx\ histogram corresponding to the test sample's \dedx, and count the number of entries in that bin in the expected \dedx\ histogram ($N$). The probability in plane (transverse cell) $i$ is calculated as $P_i=\frac{N\times n_{bin}}{N_{tot}},$ where $n_{bin}$ is the number of bins in the \dedx\ histogram, and $N_{tot}$ is the total number of entries in that histogram. The scale factor of $n_{bin}$ is applied in order to avoid issues with the limit of machine precision when we sum $\ln(P_i)$ over planes or transverse cell indices to calculate the likelihood. Figure~\ref{fig:evtlongdedx19578} shows the measured plane \dedx\ of a signal-like event in FD superposed on the probability density as a function of \dedx\ in each plane created by electrons in the FD signal neutrino MC.

\begin{figure}[h]
\begin{center}
\includegraphics[width=12cm]{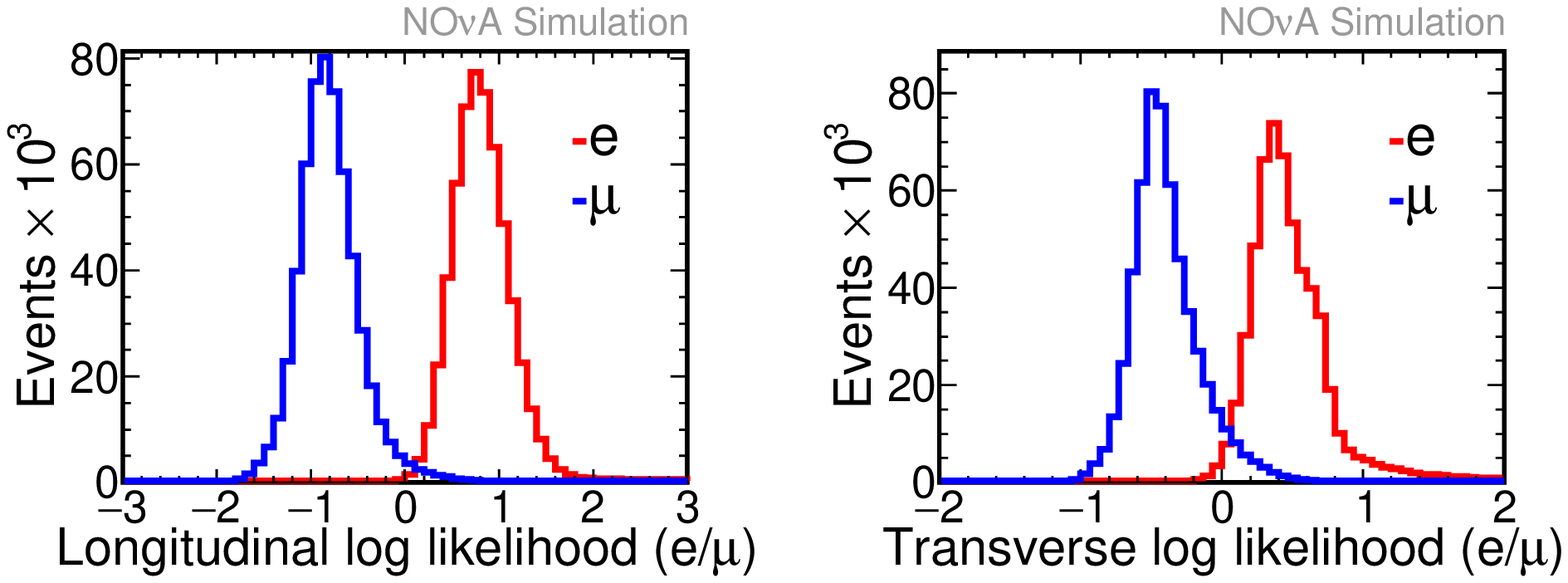}
\caption{(left) Longitudinal $e^{-}$ likelihood minus $\mu$ likelihood for $e^{-}$ (red) and $\mu$ (blue); (right) Transverse $e^{-}$ likelihood minus $\mu$ likelihood for $e^{-}$ (red) and $\mu$ (blue).}\label{fig:emull}

\includegraphics[width=12cm]{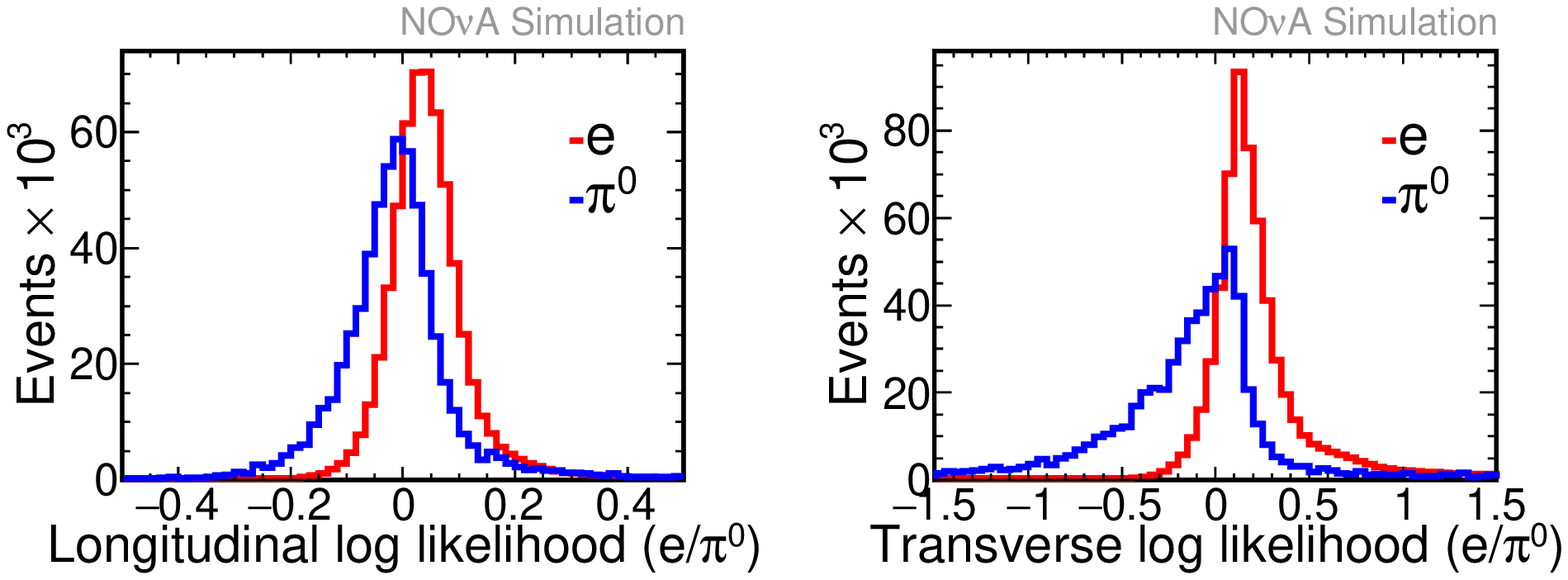}
\caption{(left) Longitudinal $e^{-}$ likelihood minus $\pi^0$ likelihood for $e^{-}$ (red) and $\pi^0$ (blue); (right) Transverse $e^{-}$ likelihood minus $\pi^0$ likelihood for $e^{-}$ (red) and $\pi^0$ (blue).}\label{fig:epi0ll}
\end{center}
\end{figure}

The likelihood of each particle's hypothesis in the $i$-th plane (transverse cell index) is defined as: $LL_i = \ln(P_{i}).$ The overall longitudinal and transverse log likelihoods are defined as: $LL_L = \Sigma LL_i/N_{p} $ and $LL_T = \Sigma LL_i/N_{t},$ where $N_{p}$ is the number of planes and $N_{t}$ is the number of transverse cells in the shower. Differences between longitudinal and transverse log likelihoods for the electron and other particle hypotheses can be used to identify electrons. Differences between longitudinal and transverse log likelihoods for electron and other particle hypotheses can be used to identify electrons. As examples, Figures~\ref{fig:emull} and \ref{fig:epi0ll} show the difference of the electron likelihood and the likelihood of the $\mu/\pi^0$ hypotheses in Monte Carlo neutrino events.  These distributions illustrate the discrimination power against major backgrounds.

\subsection{\bf Input variables and training of the artificial neural network} \label{dedxlikelihood}

Seventeen variables are used to form the input of an artificial feed-forward neutral network (ANN) to produce the likelihood-based $\nu_e$ identification variable - LID.  The first category of inputs consists of 12 differences between longitudinal and transverse log likelihoods for electron and the six other-particle hypotheses ($\gamma$, $\mu$, $\pi^0$, $p$, $n$ and changed $\pi$) applied to the most energetic shower [$LL_L(e) - LL_L(\gamma)$, $LL_T(e) - LL_T(\gamma)$, etc]. In addition to these log likelihoods, 5 additional inputs are used: (13) $\pi^0$ mass, the invariant mass of the most energetic shower with all the other showers is the slice is computed and the one closest to the $\pi^0$ mass is recorded to help reject $\pi^0$Õs in neutral current interactions; (14) shower energy fraction, shower energy divided by the total event energy; (15) vertex energy, the calorimetric energy, excluding the leading shower, within $\pm8$ planes of the event vertex;  (16) gap, the distance of the start point of the shower from the event vertex, which is small for electrons and large for photons; (17) $\cos\theta$, angle of the leading shower with respect to the beam direction, for rejection of neutral-current interactions and cosmic rays.

% (1) $LL_L(e/\gamma) = LL_L(e) - LL_L(\gamma)$; (2) $LL_T(e/\gamma) = LL_T(e) - LL_T(\gamma)$; (3) $LL_L(e/\mu) = LL_L(e) - LL_L(\mu)$; (4) $LL_T(e/\mu) = LL_T(e) - LL_T(\mu)$; (5) $LL_L(e/\pi^0) = LL_L(e) - LL_L(\pi^0)$; (6) $LL_T(e/\pi^0) = LL_T(e) - LL_T(\pi^0)$; (7) $LL_L(e/p) = LL_L(e) - LL_L(p)$; (8) $LL_T(e/p) = LL_T(e) - LL_T(p)$; (9) $LL_L(e/n) = LL_L(e) - LL_L(n)$; (10) $LL_T(e/n) = LL_T(e) - LL_T(n)$; (11) $LL_L(e/\pi) = LL_L(e) - LL_L(\pi)$; (12) $LL_T(e/\pi) = LL_T(e) - LL_T(\pi)$.

The neural network has one input layer, three hidden layers and one output layer. The input layer has 17 nodes which correspond to the 17 input variables. The output layer has one node. The optimal architecture of the three hidden layers was determined to be 22:12:6.

\subsection{\bf Performance and choice of PID}
\begin{figure}[h]
 \begin{center}
    \includegraphics*[width=10cm]{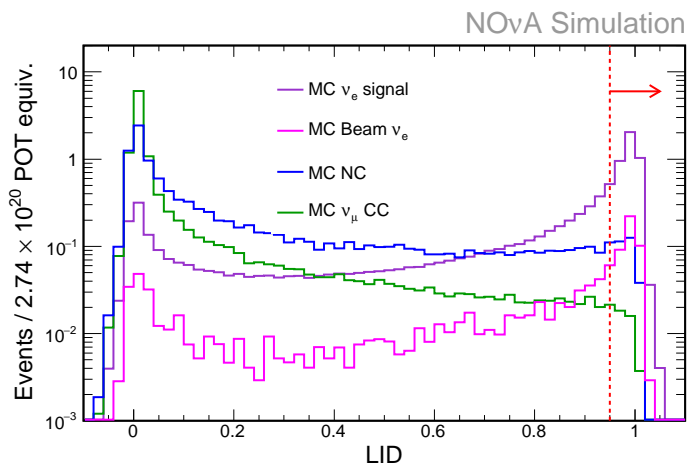}
  \caption{LID on FD neutrino MC (log. scale).}
\label{fig:lid}
 \end{center}
  \end{figure}

The output of the artificial neutral network (LID) can be found in Figure~\ref{fig:lid}. The dashed line and arrow show the cut applied to the first $\nu_e$ analysis. The value, LID$>0.95$, is obtained by maximizing the figure of merit defined as $FOM=S/\sqrt{B}$, where $S$ is the number of signal and $B$ is the number of the background. The intention of using $FOM=S/\sqrt{B}$ instead of $FOM=S/\sqrt{S+B}$ is to optimize the ability to reject the zero $\nu_\mu\to\nu_e$ hypothesis. The $\nu_e$ selection efficiency achieves an efficiency of $34\%$ relative to the contained sample, while rejecting more than $99\%$ of beam backgrounds.

Similarly to LID, the cut for selecting $\nu_e$ events with LEM is determined to be LEM$>0.8$. The $\nu_e$ selection efficiency is $36\%$ relative to the contained sample, while rejecting more than $99\%$ of beam backgrounds. There is $62\%$ overlap of selected signal events between LID and LEM.Ê

Both EIDs are very similar in signal efficiency, figure of merit (FOM), systematic uncertainties, and overall sensitivity to $\nu_e$ appearance and oscillation parameters. Prior to unblinding we decided to present results of both selections and to use the LID technique as the primary result.

%After fully considering the ease of explanation, cosmic background rejection, uncertainties in signal and background efficiencies and effects in calibration,

\section{\bf Event selection and FD prediction}

A series of data quality cuts are applied to ensure that the data subjected to the full analysis were taken under the normal beam and detector conditions. Events in a 12-$\mu$s window around the 10-$\mu$s neutrino spill time are selected to reject cosmic rays collected in the 550-$\mu$s trigger window. Prior to the PID selection, reconstruction quality and containment cuts are applied to remove particles that are poorly reconstructed or that enter from the edges of the detector. An additional cut on the ratio of the transverse momentum to the total momentum of the event ($P_T/P$) is applied to require the orientation of showers to match the beam direction. The oscillation maximum is around 2 GeV. To further improve the figure of merit, the slice calorimetric energy $E$ is required to be 1.5 GeV $< E <$ 2.7 GeV (1.3 GeV $< E <$ 2.7 GeV for LEM). Finally, the LID value is required to be greater than 0.95 (0.8 for LEM) to select $\nu_e$ CC candidates. After all selection cuts, both EIDs have an overall rejection rate of 1 in $10^8$ for cosmogenic backgrounds. Based on the cosmic ray data, we predict 0.06 cosmic background events for both LID and LEM \cite{ref:evtsel}. The consistency between data and MC has been validated based on the near detector data and cosmic ray data \cite{ref:evtsel, Duyang:2015cvk, Sachdev:2015hpa, ref:syst}.

%This event selection achieves 350 million-to-1 cosmic background rejection rate.

The Near Detector data provide a data-driven correction for the MC normalization in the Far Detector. We scale up each background component ($\nu_\mu$ CC, NC and beam $\nu_e$ CC) in the Far Detector MC by the extrapolated data/MC ratio in the Near Detector to make the background prediction. Because there are no oscillated $\nu_e$ CC events in the Near Detector, we correct the Far Detector $\nu_e$ CC signal by the extrapolated data/MC difference of selected  $\nu_\mu$ CC events in the Near Detector. The Near-to-Far extrapolation ratios for signal and backgrounds are determined with Monte Carlo samples. This extrapolation cancels most of the systematic errors. Remaining systematic uncertainties after the extrapolation are evaluated by extrapolating ND data with nominal MC and systematically modified MC samples, with variations to the calibration, non-linearity in detector responses, the neutrino-interaction model, the ND background composition, overall normalization, beam uncertainties, and other smaller uncertainties. Summed in quadrature the total uncertainty is $17.6\%$ for the signal and $10.1\%$ for the background under the LID selection~\cite{Sachdev:2015hpa, ref:syst}.

%The full-detector equivalent exposure used in the first analysis is $2.74\times10^{20}$ POT.
Based on the Near Detector data and cosmic ray data, the expected background in FD is 0.94$\pm$0.10 (syst.) events for the LID selector. Oscillation parameters that produce this prediction are $\sin^2\theta_{23} = 0.5$, $\Delta m^2_{32} = +2.37\times10^{-3}$eV$^{2}$, $\sin^2 2\theta_{12} = 0.846$, $\Delta m^2_{21} = 7.53\times10^{-5}$eV$^2$, $\sin^{2}2\theta_{13} = 0.086$, and $\delta_{CP} = 0$. The background prediction varies about $\%1$ for different choices of oscillation parameters. The dominant backgrounds are $49\%$ beam $\nu_e$ CC and  $38\%$ NC. $\nu_\mu$ CC and cosmic rays are each about $6\%$.  For the LEM selector, we predict to have 1.00$\pm$0.12 (syst.) background events with  $46\%$ beam $\nu_e$ CC and  $40\%$ NC in them. $\nu_\mu$ CC and cosmic rays are at the same level as LID.

The signal prediction depends on the choice of oscillation parameters. The highest prediction is made under the assumption of normal mass hierarchy, $\delta_{CP}$ close to $3\pi/2$ and   $\theta_{23}=\pi/4$, where we expect 5.62$\pm$0.99 $\nu_{\mu}\to\nu_e$ oscillation events selected by LID and 5.91$\pm$0.89 $\nu_{\mu}\to\nu_e$ oscillation events selected by LEM. Before unblinding, we checked sidebands of EIDs and energy. Results of both are consistent with expectations.

\begin{figure}[h]
\begin{center}
\begin{minipage}[t]{12cm}
\includegraphics[width=6cm]{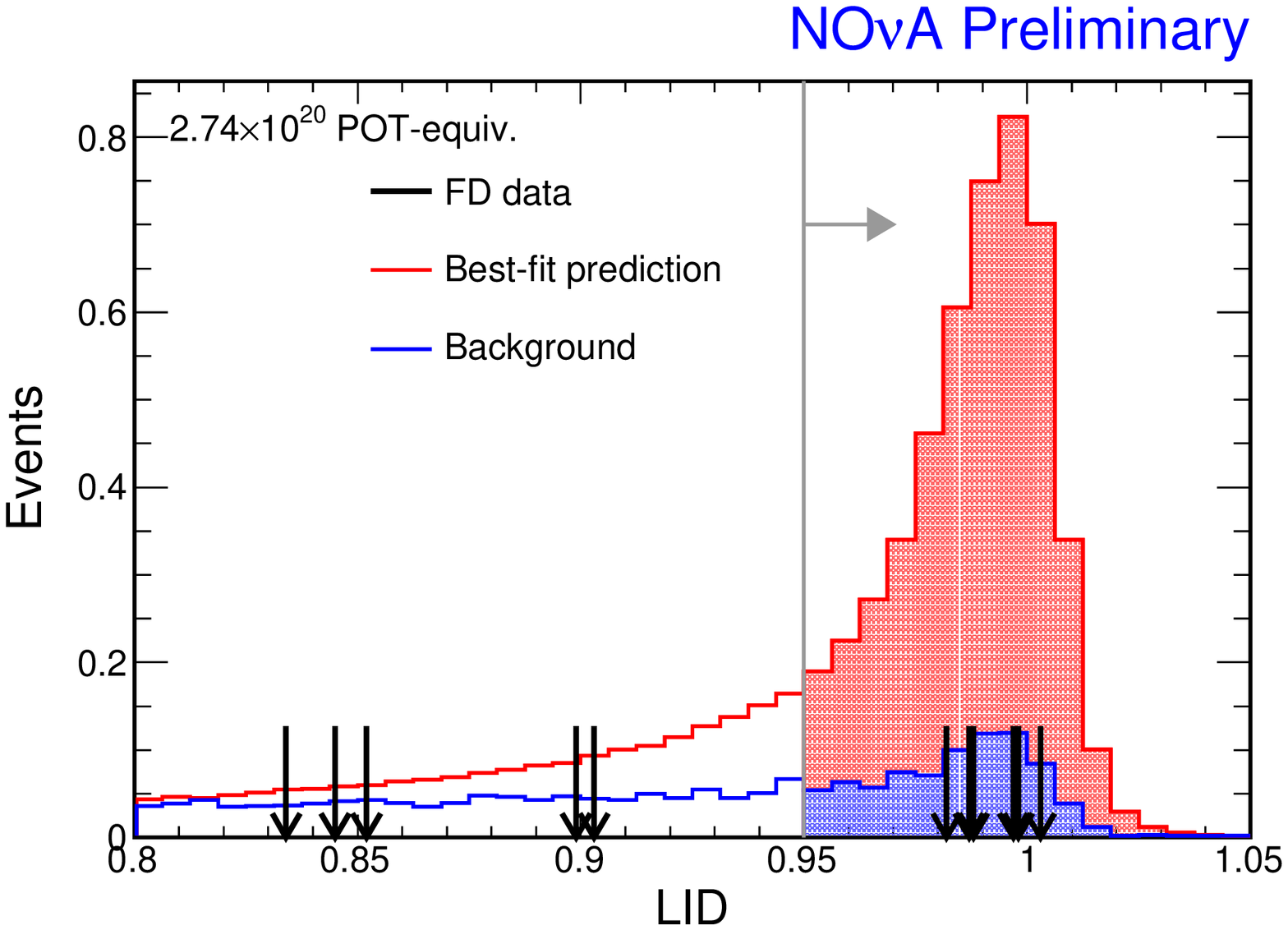}
\includegraphics[width=6cm]{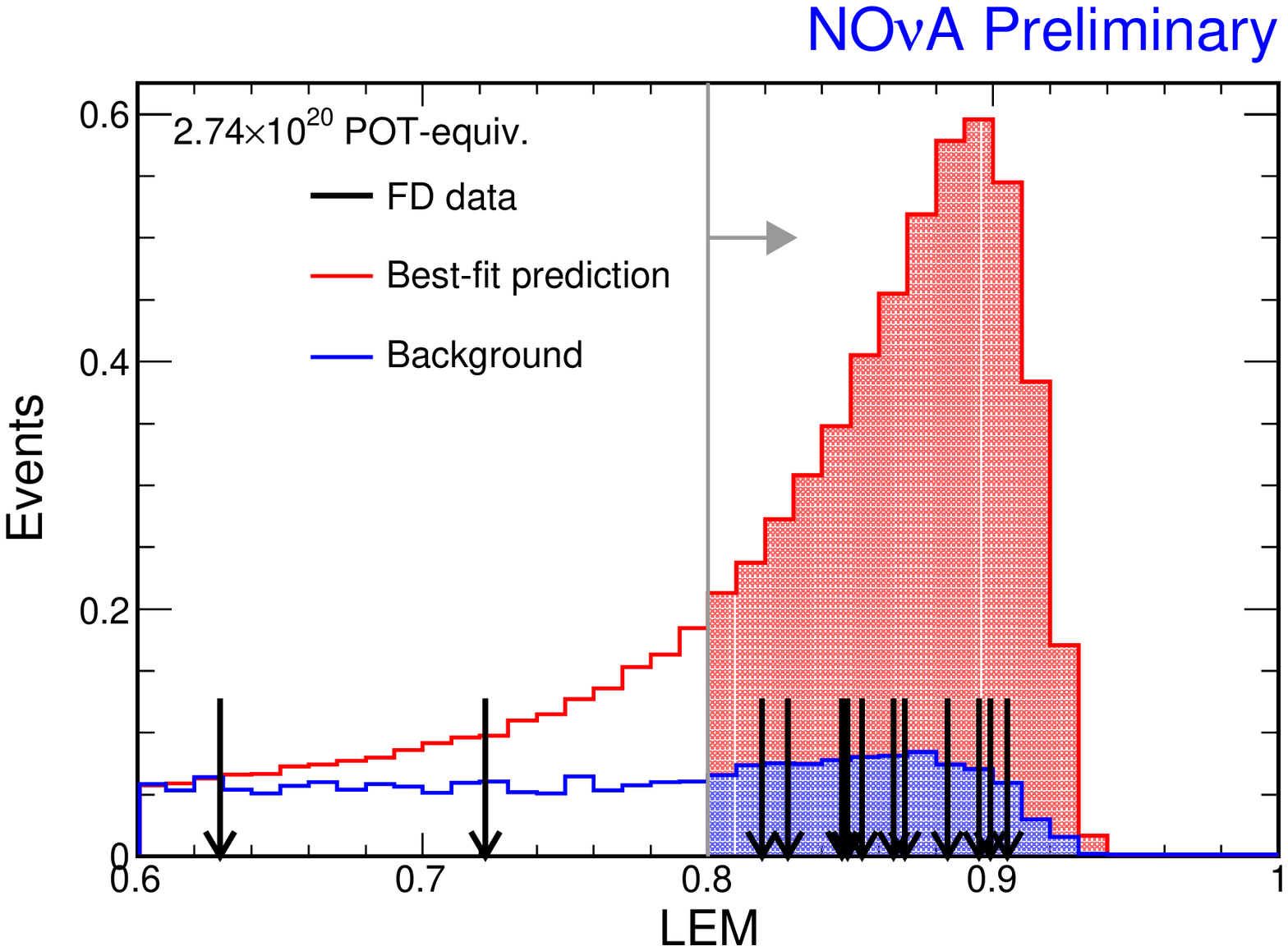}
\end{minipage}
\caption{PID distributions for FD data and MC: (left) LID (right) LEM. Gray lines and arrows are PID cuts and selected signal  region}
\label{fig:fdpid}
\end{center}
\end{figure}

\section{\bf Results}
In the Far Detector data, 6 events are selected by LID$>$0.95 and 11 events are selected by LEM$>$0.8, as shown in Figure~\ref{fig:fdpid}.  In the data the (LID only)/(LEM only)/(LID and LEM) events are 0/5/6. The 5 events that are only selected by LEM have LID values in the range $0.7-0.95$, which is just below the signal region LID$>0.95$.  As shown in Figure~\ref{fig:fdpid} (left), in this LID region the probability of signal (red line) is still higher than background (blue line). In addition, 2 of the 5 LEM-only events also fail the energy cut $1.5-2.7$ GeV for LID. Given the expected correlations, (LID only)/(LEM only)/(LID and LEM) events in MC signal and backgrounds, the observed event counts yield a reasonable mutual p-value of $10\%$.

%Considering that our current event selection is optimized by maximizing $S/\sqrt{B}$, which is appropriate for rejecting zero-oscillation hypothesis, but is less so for the oscillation parameter measurements, we expect to have large statistical fluctuations on the selected events.

\begin{figure}[h]
\begin{center}
\begin{minipage}[t]{12cm}
\includegraphics[width=6cm]{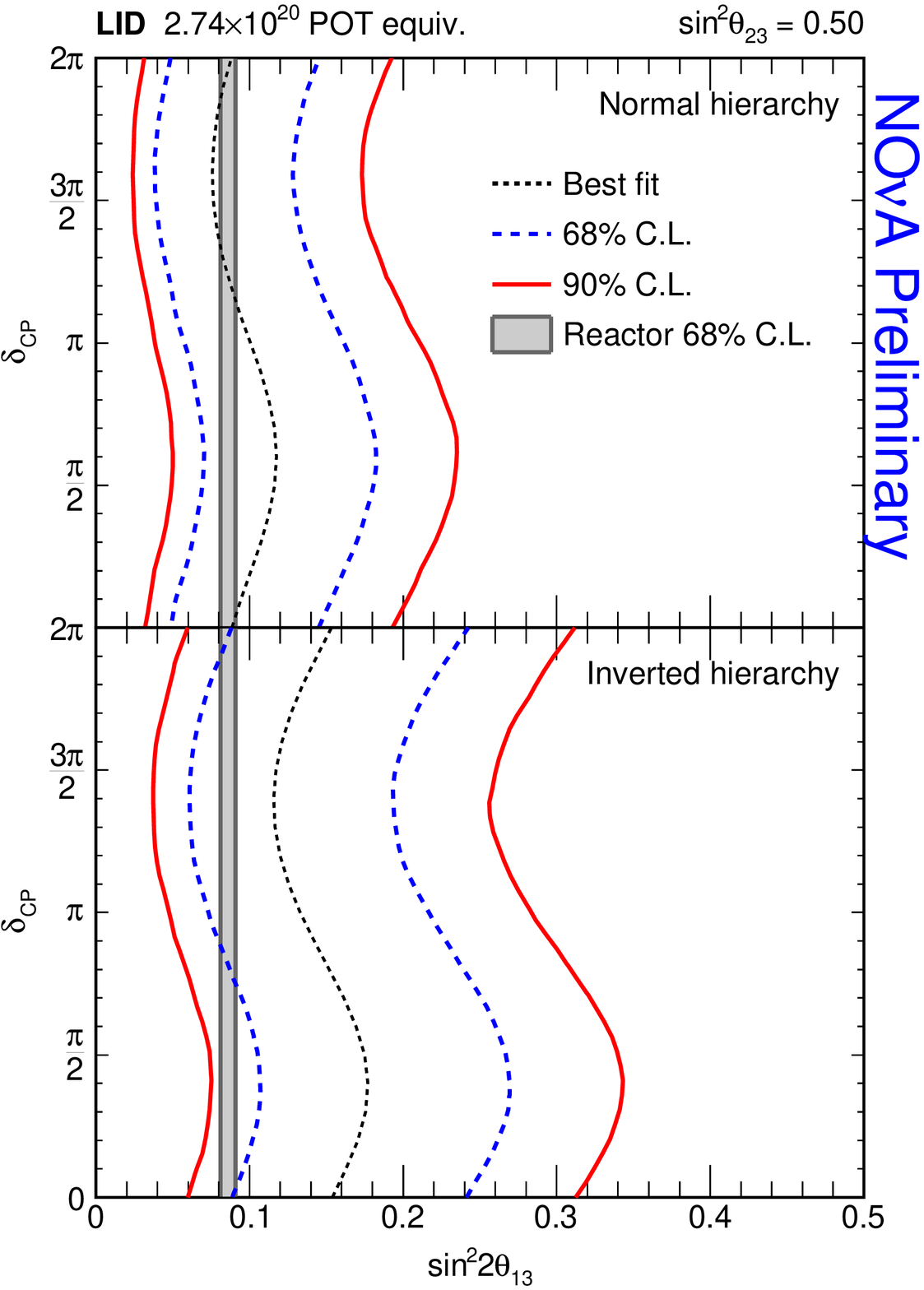}
\includegraphics[width=6cm]{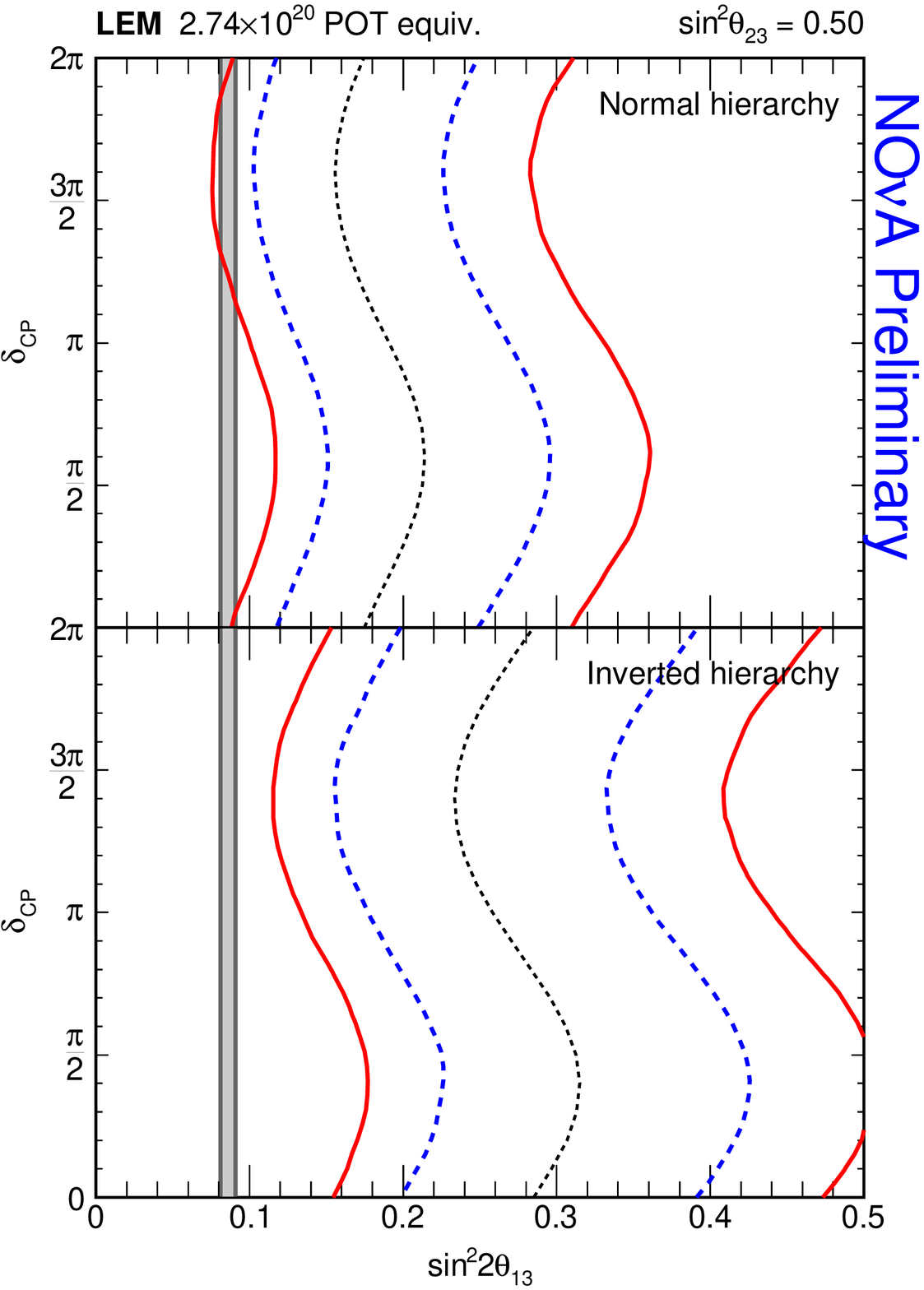}
\end{minipage}
\caption{The allowed regions of $\delta_{CP}$ and $\sin^2 2\theta_{13}$ with the LID (left) and LEM (right) analysis. World average reactor $\sin^2 2\theta_{13}$ results are drawn as the gray bands.}
\label{fig:conts}
\end{center}
\end{figure}

The $\nu_e$ appearance significance based on the 6 $\nu_e$ candidates selected by LID is 3.3$\sigma$ (5.4$\sigma$ for LEM). Allowed regions of $\delta_{CP}$ and $\sin^2 2\theta_{13}$ are obtained by the Feldman-Cousins (FC) approach to interpret the measured $-2\Delta log(L)$ in the data~\cite{Feldman:1997qc}. When generating pseudo-experiments at each ($\delta_{CP}$, $\sin^2 2\theta_{13}$) point, central values and variations of other oscillation parameters are listed below: $\Delta m^2_{21}=7.53\pm 0.18 \times 10^{-5}$eV$^2$ and $\sin^2 2\theta_{12}=0.846\pm0.021$ are taken from PDG \cite{Agashe:2014kda}, $\Delta m^2_{32}=2.37\pm0.15\times10^{-3}$ eV$^2$ [normal mass hierarchy (NH)], $-2.40\pm0.15\times10^{-3}$ eV$^2$ [inverted mass hierarchy (IH)] is taken from the first NOvA $\nu_\mu$ disappearance measurement that is parallel to this analysis, $\sin^2\theta_{23}$ is held at 0.5. The resulting allowed regions produced by LID and LEM are shown in Figure~\ref{fig:conts}. World average reactor result $\sin^2 2\theta_{13}=0.086\pm0.05$ are also shown as grey bands. For LID, the $68\%$ C.L. interval for the NH hypothesis is compatible with the reactor result for all $\delta_{CP}$ values.

\begin{figure}[h]
\begin{center}
\begin{minipage}[t]{12cm}
\includegraphics[width=6cm]{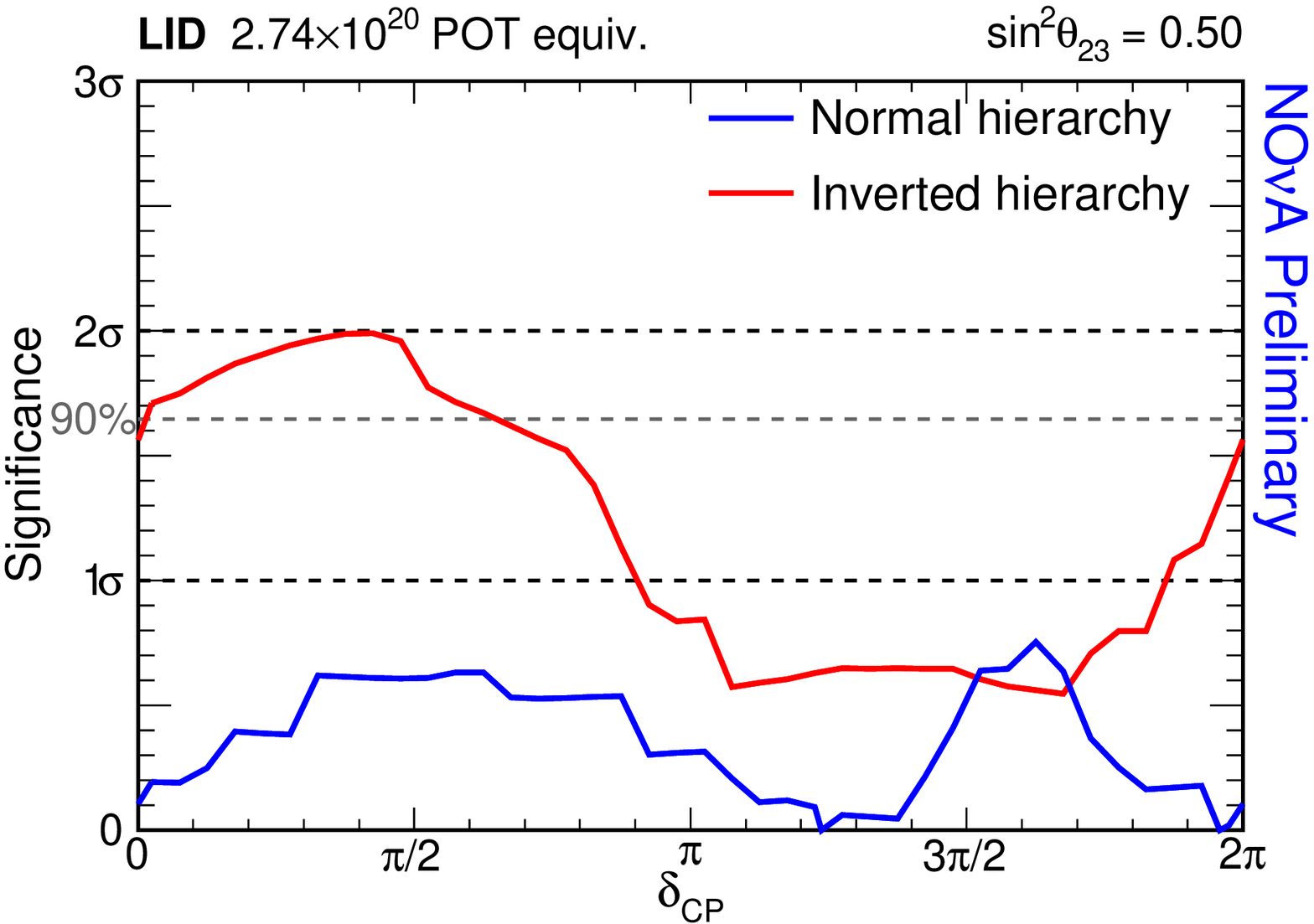}
\includegraphics[width=6cm]{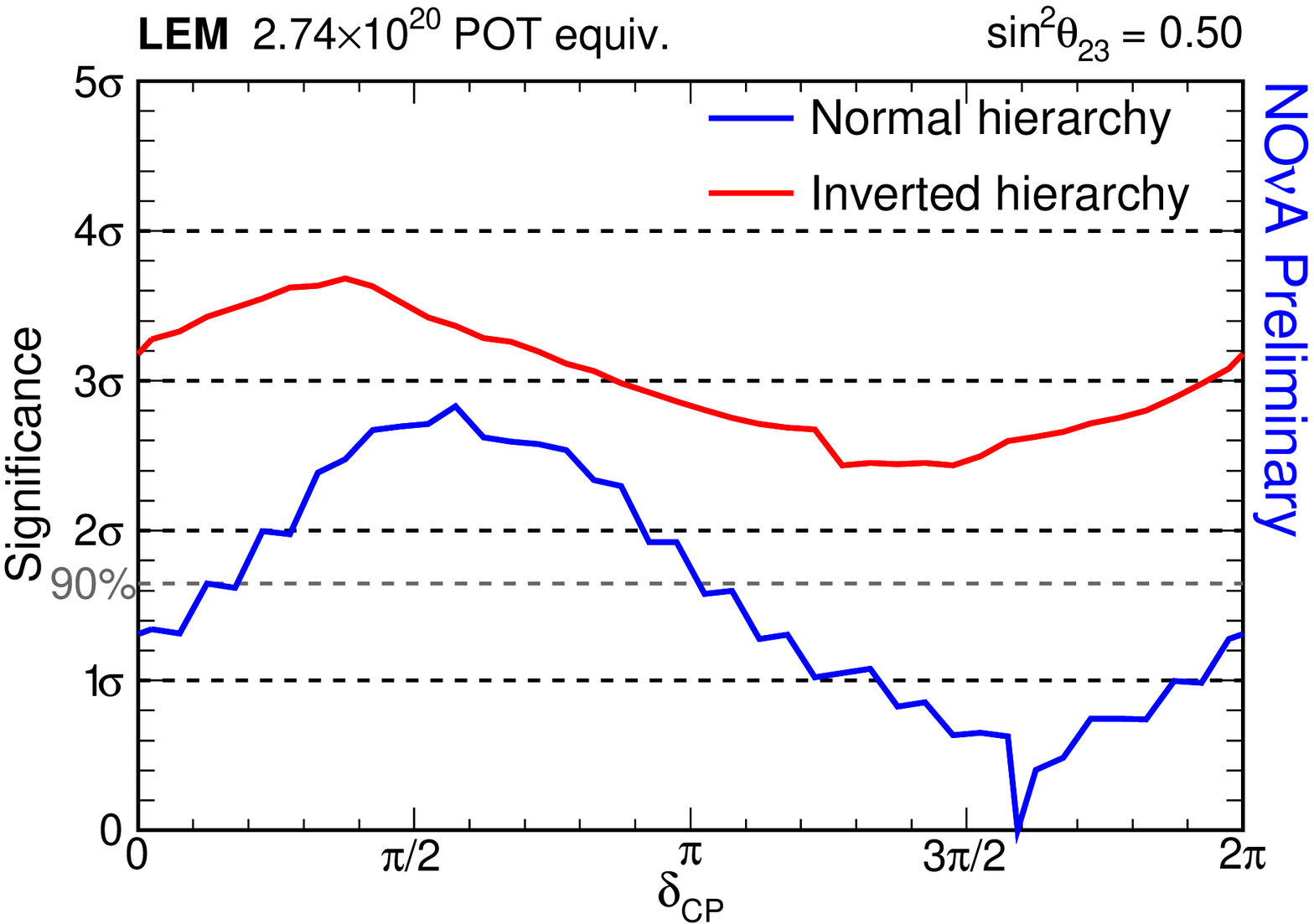}
\end{minipage}
\caption{Significance as a function of $\delta_{CP}$: (left) LID (right) LEM. }
\label{fig:fddcp}
\end{center}
\end{figure}

By varying $\sin^2 2\theta_{13}$ values within the global reactor constraint, we converted the 2-D FC contours into significance as a function of $\delta_{CP}$, as shown in Figure~\ref{fig:fddcp}. The LID result [Figure~\ref{fig:fddcp} (left)] shows that the range of $0 < \delta_{CP} < 0.65\pi$ in the IH is disfavored at the 90$\%$ C.L. in the case of $\sin^2\theta_{23}=0.5$. The number of events selected by LEM is greater than that predicted even in the NH, $\delta_{CP} = 3/2\pi$ case, but $12\%$ of pseudo-experiments generated in this scenario have an equal or worse $\chi^2$ than that observed in the data. The LEM result [Figure~\ref{fig:fddcp}(left)] shows that IH is disfavored at $>2.2\sigma$ while NH for $0<\delta_{CP}<\pi$ is mildly disfavored ($>1\sigma$). Both the LID and LEM results prefer normal ordering in most of the $\delta_{CP}$ range, and prefer $\delta_{CP}$ near $3\pi/2$.

%NH is favored within $1\sigma$. We also varied $\sin^2\theta_{23}$ value from 0.4 to 0.6 and found that IH for $0<\delta_{CP}<0.8\pi$ is mildly disfavored by $1\sigma$ in all cases for LID.

\end{document}

%% file: econfmacros.tex
\def\beq{\begin{equation}}
\def\eeq#1{\label{#1}\end{equation}}
\def\eeqn{\end{equation}}

\def\beqa{\begin{eqnarray}}
\def\eeqa#1{\label{#1}\end{eqnarray}}
\def\eeqan{\end{eqnarray}}

\let\bar=\overbar

\def\Dslash{\not{\hbox{\kern-4pt $D$}}}
\def\dslash{\not{\hbox{\kern-2pt $\del$}}}

\def\msb{{\bar{\ssstyle M \kern -1pt S}}}